\documentclass[%
prl,aps,amsmath,superscriptaddress,showpacs,twocolumn%
]{revtex4}
\usepackage{bm}
\usepackage{amsmath}
\usepackage{epsfig}
\begin{document}
\title{Anisotropic admixture in color-superconducting quark matter}
\author{Michael Buballa}
\affiliation{Institut f{\"u}r Kernphysik, Schlossgartenstr.\ 9,
D-64289 Darmstadt, Germany}
\affiliation{Gesellschaft f{\"u}r Schwerionenforschung, Planckstr.\ 1,
D-64291 Darmstadt, Germany}
\author{Ji\v r\'{\i} Ho\v sek}
\affiliation{Dept. Theoretical Physics, Nuclear Physics Institute,
             25068 \v Re\v z (Prague), Czech Republic }
\author{Micaela Oertel} 
\affiliation{IPN-Lyon, 43 Bd du 11 Novembre 1918,
             69622 Villeurbanne C\'edex, France}

\date{April 2, 2003}
\begin{abstract}
The analysis of color-superconducting two-flavor deconfined quark matter
at moderate densities is extended to include a particular spin-1 Cooper
pairing of those quarks which do not participate in the standard spin-0
diquark condensate. 
(i) The relativistic spin-1 gap $\Delta'$ implies spontaneous breakdown of
rotation invariance manifested in the form of the quasi-fermion dispersion
law.
(ii) The critical temperature of the anisotropic component is approximately
given by the relation $T_c'\simeq \Delta'(T=0)/3$. 
(iii) For massless fermions the gas of anisotropic Bogolyubov-Valatin
quasiquarks becomes effectively gapless and two-dimensional. Consequently,
its specific heat depends quadratically on temperature. 
(iv) All collective Nambu-Goldstone excitations of the anisotropic phase
have a linear dispersion law and the whole system remains a superfluid.
(v) The system exhibits an electromagnetic Meissner effect.
\end{abstract}
\maketitle
Recent investigations suggest that the phase structure of QCD is very rich
\cite{RaWi00,Al01}.  At low temperatures and high densities strongly
interacting matter is expected to be a color superconductor \cite{CoPe75}.  At
asymptotically high densities, where the QCD coupling constant becomes small,
this can be analyzed starting from first principles
\cite{Son99,PiRi}, whereas at more moderate densities, present
(presumably) in the interiors of neutron stars, these methods are no
longer justified. In this region the low-energy dynamics of deconfined quark 
matter is often studied employing effective Lagrangians
${\cal L}_{\mathit{eff}}$ which contain local or non-local four-fermion 
interactions, most importantly interactions derived from 
instantons or on a more phenomenological basis \cite{RSSV,ARW98,models}.
The non-confining gluon $SU(3)_c$ gauge fields are then treated as weak
external perturbations, and neglected in lowest approximation.

In this letter we consider the case of two flavors
which is most likely relevant at chemical potentials just above the
deconfinement phase transition. On physical grounds it is then natural to
assume that ${\cal L}_{\mathit{eff}}$ favors the spontaneous formation of
spin-0 isospin singlet Cooper pair condensates \cite{BaLo84,RaWi00}
$
 \delta \;=\;\langle \psi^T \;C\,\gamma_5\,\tau_2\;\lambda_2\; \psi \rangle~,
$
where $\psi$ is a quark field, $C$ the matrix of charge conjugation,
$\tau_2$ a Pauli matrix which acts in flavor space, and $\lambda_2$ a
Gell-Mann matrix which acts in color space. 
Due to the latter $SU(3)_c$ is broken down to $SU(2)_c$.
This has the following consequences
for the physical excitations of the system: 

(i) Corresponding to the mixing of the colors 1 and 2 
there are two Bogolyubov-Valatin quasiquarks for each flavor
with the dispersion law
$
E_{1}^{\pm}(\vec p) = E_2^{\pm}(\vec p) \equiv E^{\pm}(\vec p) 
= \sqrt{(\epsilon_p\pm\mu)^2 + |\Delta|^2}~.
\label{E1}
$
The energy gap $\Delta$ is the solution of a selfconsistent gap equation
and is found to be typically of the order $\sim 100$~MeV in model
calculations \cite{ARW98,RSSV,models}.
$\epsilon_p = \sqrt{\vec p^{\,2} + M^2}$, where $M$ is an 
effective Dirac mass, related to the chiral condensate 
$\langle\bar\psi\psi\rangle$ via a selfconsistency 
equation~\cite{models}. 
For each flavor there is an unpaired quark of color 3 
with the dispersion law 
$\epsilon_3^{\pm}(\vec p) \,=\, \epsilon_p \,\pm\,\mu$.

(ii) Because of the spontaneous breaking of $SU(3)_c$ down to $SU(2)_c$
five of the eight gluons receive a mass (Meissner effect), whereas three 
remain massless \cite{RiCD00}.
Since no global symmetry is spontaneously broken 
there are no massless Goldstone bosons.

(iii) The condensate $\delta$ is invariant under a local $U(1)$ transformation
generated by $\tilde Q = Q - \frac{1}{2\sqrt{3}} \lambda_8$, where $Q$
is the electromagnetic charge operator and $\lambda_8$ a Gell-Mann matrix
in color space. As long as this symmetry is not broken by other condensates, 
there is a ``new'' photon (a linear combination of the ``normal'' photon
and the eighth gluon) which remains massless. This means, 
there is no electromagnetic Meissner effect.
 
According to Cooper's theorem any attractive interaction leads to an 
instability at the Fermi surface.
It is therefore rather unlikely, that the Fermi sea of color-3 quarks
stays intact. As only quarks of a single color are involved, 
the pairing must take place in a channel which
is symmetric in color. Assuming $s$-wave condensation in an 
isospin-singlet channel, a possible candidate is a spin-1 condensate
~\cite{ARW98}.
Although the size of the corresponding gap was estimated to be much
smaller than $\Delta$~\cite{ARW98}, its existence can have important
astrophysical consequences.  For example,
if all quarks are gapped, 
the specific heat of a potential quark core of a neutron
star (and hence the cooling of the star) is goverened by the 
size of the {\it smallest} gap~\cite{RaWi00}.
The same is true for other transport properties, like neutrino
emissivity or viscosity.

This letter is devoted to a quantitative analysis of this 
possibility. To this end we consider the condensate
\begin{equation}
  \delta' \;=\; \langle \psi^T \;C\,\sigma^{03}\;\tau_2\;\hat P_3^{(c)}\;\psi 
  \rangle~,
\label{qqt}
\end{equation}
where $\sigma^{\mu\nu} = i/2\,[\gamma^\mu,\gamma^\nu]$ and
$\hat P_3^{(c)} = 1/3 - 1/\sqrt{3}\,\lambda_8$ is 
the projector on color 3. 
$\delta'$ 
is a 
ground-state expectation value of a complex vector order parameter
$\phi^{0n}\equiv\phi_n$ describing spin 1 and breaking spontaneously the 
rotational invariance of the system.
There are well-known examples for spin-1 pairing in condensed matter physics,
e.g., superfluid $^3He$, where some phases are also anisotropic~\cite{He3}.  
In relativistic systems this is  certainly not a very frequent phenomenon.
It is possible only at finite chemical potential, which itself breaks 
Lorentz invariance explicitly. (Relativistic Cooper pairing into
spin-1 with nonzero angular momentum was considered elsewhere, e.g.,
\cite{BaLo84,Sch00}.)
Another example of non-isotropic quark pairing are crystalline
color superconductors~\cite{crystal}.
The role of spin-1 condensates as an alternative to crystals
in single color or single flavor systems has also been discussed in
Ref.~\cite{ABCC02}.

Since rotational invariance is a global symmetry of the primary Lagrangian,
an arbitrary small gap of the anisotropic phase implies specific
gapless collective excitations with given Landau critical velocity
crucial for the superfluid behavior of the system. We will briefly discuss
this at the end of this Letter.
It is also interesting to note that $\delta'$ is
not neutral with respect to the ``rotated'' electric charge $\tilde Q$ 
and there is no generalized electric charge for which both,
$\delta$ and $\delta'$, are neutral. This means, if both, $\delta$ and 
$\delta'$, were present in a neutron star, there would be an electromagnetic 
Meissner effect, which would strongly influence the magnetic field.
Recently, similar effects have been discussed in Ref.~\cite{SWR03}.
The detailed evaluation of the Meissner masses for our case remains 
to be done.   

For the quantitative analysis we have to specify the interaction.
Guided by the structure of instanton-induced interactions 
(see, e.g., \cite{RSSV}) we consider a quark-antiquark term
\begin{equation}
{\cal L}_{q\bar q} = G \Big\{ (\bar \psi\psi)^2 - (\bar \psi\vec\tau\psi)^2
   - (\bar\psi i\gamma_5\psi)^2 + (\bar \psi i\gamma_5\vec\tau\psi)^2 \Big\}
\label{Lqqbar}
\end{equation}
and a quark-quark term
\begin{alignat}{1}
{\cal L}_{qq}\;=\;&-H_s \sum_{{\cal O} = \gamma_5, 1}
                      (\bar \psi {\cal O} C\tau_2\lambda_A\bar\psi^T)
                      (\psi^T C {\cal O} \tau_2\lambda_A\psi^T)
\nonumber \\
                 &-H_t (\bar \psi \sigma^{\mu\nu} C\tau_2\lambda_S\bar\psi^T)
                      (\psi^T C \sigma_{\mu\nu} \tau_2\lambda_S\psi^T)~,
\label{Lqq}
\end{alignat}
where $\lambda_A$ and $\lambda_S$ are the antisymmetric and symmetric
color generators, respectively. For instanton induced interactions
the coupling constants fulfill the relation
$G : H_s : H_t =  1 : \frac{3}{4} : \frac{3}{16}$,
but for the moment we will treat them as arbitrary parameters.
As long as they stay positive, the interaction is attractive in the
channels giving rise to the diquark condensates $\delta$ and 
$\delta'$ as well as to the chiral condensate 
$\langle\bar\psi\psi\rangle$.
It is straight forward to calculate the mean-field 
thermodynamic potential $\Omega(T,\mu)$ in the presence of these condensates:
\begin{alignat}{1}
    \Omega(T,\mu) = &-4 \sum_{i=1}^3 \sum_{+-} \int\!\!\frac{d^3p}{(2\pi)^3}
                        \Big[\frac{E_i^\pm}{2} 
    + T\ln\Big(1+e^{-E_i^\pm/T}\Big) \Big]
\nonumber\\
&\hspace{-.2cm}+\frac{1}{4G}(M-m)^2 +\frac{1}{4H_s}|\Delta|^2
+\frac{1}{16H_t}|\Delta'|^2~,
\label{Omega}
\end{alignat}
where $m$ is the bare quark mass, $M = m -2G\langle\bar\psi\psi\rangle$,
$\Delta= -2H_s\delta$, and $\Delta' = 4H_t\delta'$.
The dispersion law for quarks of color 3 reads
\begin{equation}
E_3^\mp(\vec p) \;=\; \sqrt{ (\sqrt{M_{\mathit{eff}}^2 + \vec p^{\,2}} \mp \mu_{\mathit{eff}})^2
                            + |\Delta_{\mathit{eff}}'|^2 }~,
\label{E3}
\end{equation}  
where 
$\mu_{\mathit{eff}}^2 = \mu^2 + |\Delta'|^2 \sin^2{\theta}$,
$M_{\mathit{eff}} = M \mu/\mu_{\mathit{eff}}$, and
$
|\Delta_{\mathit{eff}}'|^2 = |\Delta'|^2\,(\cos^2{\theta} + M^2/\mu_{\mathit{eff}}^2\,\sin^2{\theta})~. 
$
Here $\cos{\theta} = p_3/|\vec p|$. 
Thus, as expected, for $\Delta'\neq 0$, $E_3^\pm(\vec p)$ is an anisotropic function of $\vec p$, 
clearly exhibiting the spontaneous breakdown of rotational invariance. 
For $M = 0$, the gap $\Delta_{\mathit{eff}}'$ vanishes at $\theta = \pi/2$. 
In general its minimal value is given by
$ \Delta'_0 = M |\Delta'|/\sqrt{\mu^2 + |\Delta'|^2}$~.
Expanding $E_3^-$ around its minimum the low-lying quasiparticle
spectrum takes the form
\begin{equation}
  E_3^-(p_\perp, p_3) \approx \sqrt{ \Delta_0^{\prime 2} + v_\perp^2 (p_\perp - p_0)^2
                                  + v_3^2 p_3^2}~,  
\end{equation}
where 
$
  v_\perp = (1-(\mu {\Delta'}_0^2/(M \Delta'^2))^2)^{1/2},\;
  v_3 = \frac{\Delta'_0}{M},\; 
  p_0 = \frac{v_\perp}{v_3}|\Delta'|,
$
and $p_\perp^2 = p_1^2 + p_2^2$.
This leads to a density of states linear in energy: 
\begin{equation}
  N(E) = \frac{1}{2\pi} \frac{\mu^2 + |\Delta'|^2}{|\Delta'|}\;E\;
  \theta(E-\Delta'_0)~.
\label{NE}
\end{equation}
\begin{figure}[t]
\epsfig{file=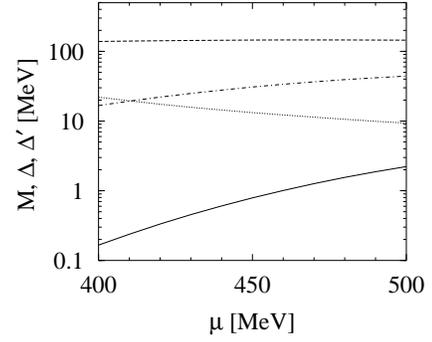, width = 5.8cm}
\caption{$M$ (dotted), $\Delta$ (dashed), and $\Delta'$ (solid) at $T$~=~0
         as functions of the quark chemical potential $\mu$ using parameter
         set 1 (see text). 
         The dashed-dotted line indicates the result for $\Delta'$
         taking parameter set 2.}
\label{cond}
\end{figure}
The actual values for $\Delta$, $\Delta'$ and $M$ follow from the
condition that the stable solutions correspond to
the absolute minimum of $\Omega$ with respect to these quantities.
Imposing $\partial\Omega/\partial{\Delta'}^* = 0$ leads to the following gap equation for $\Delta'$:
\begin{equation}
\Delta' = 16H_t\Delta' \sum_{+-}\int \frac{d^3 p}{(2\pi)^3} \, 
(1\pm\frac{{\vec p}_\perp^{\,2}}{s})\frac{1}{E_3^\pm}
\tanh{\frac{E_3^\pm}{2T}}~, 
\label{Deltapgap}
\end{equation}  
where $s = \mu_{\mathit{eff}}(\vec p^{\,2} + M_{\mathit{eff}}^2)^{1/2}$.
Similarly one can derive gap equations for $\Delta$ and $M$
by the requirements $\partial\Omega/\partial\Delta^* = 0$ and 
$\partial\Omega/\partial M = 0$, respectively. Together with 
Eq.~(\ref{Deltapgap}), they form a set of three coupled equations,
which have to be solved simultaneously.
However, the equations for $\Delta$ and $\Delta'$ are not directly 
coupled, but only through their dependence on $M$.

In our numerical calculations we use a sharp 3-momentum cutoff
$\Lambda$ to regularize the integrals.  We then have five parameters:
$m$, $\Lambda$, $G$, $H_s$, and $H_t$.  We choose $m =$~5~MeV,
$\Lambda = 600$~MeV, and $G\Lambda^2 = 2.4$ -- leading to reasonable
vacuum properties, $M = 393$~MeV and $\langle\bar u u\rangle = (-244
\rm{MeV})^3$ --, and the instanton relation to fix $H_s$ and $H_t$
("parameter set 1").  The resulting values of $M$, $\Delta$, and
$\Delta'$ as functions of $\mu$ at $T$~=~0 are displayed in
Fig.~\ref{cond}. The chemical potentials correspond to baryon
densities of about 4 - 7 times nuclear matter density.  In agreement
with earlier expectations \cite{ARW98} $\Delta'$ is small compared
with $\Delta$.  However, its value rises strongly with $\mu$.  Being a
solution of a selfconsistency problem, $\Delta'$ is also extremely
sensitive to the coupling constant $H_t$.
If we double the value of $H_t$
(``parameter set 2''), we arrive at the dashed-dotted line for 
$\Delta'$, which is then 
comparable to $\Delta$.
As a consequence of the factor $(1-{\vec p}_\perp^{\,2}/s)$ in the gap equation
(\ref{Deltapgap}), $\Delta'$ is very sensitive to value and the form
of the cutoff.

\begin{figure}[t]
\epsfig{file=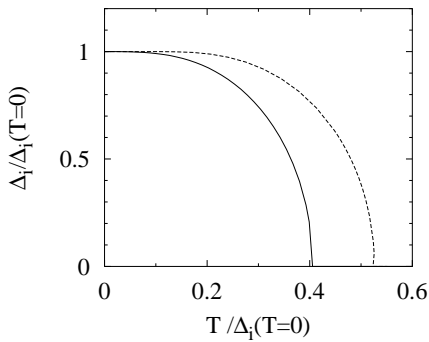,width = 5.8cm}
\caption{$\Delta_i/\Delta_i$($T$=0) as function of $T/\Delta_i$($T$=0).
Dashed: $\Delta_i=\Delta$. Solid: $\Delta_i=\Delta'$. The calculations
have been performed at $\mu=450$~MeV for parameter set 1.
}
\label{condt}
\end{figure}
With increasing
temperature both condensates, $\delta$ and $\delta'$, are reduced
and eventually vanish in second-order phase transitions at critical
temperatures $T_c$ and $T_c'$, respectively. It has been shown 
\cite{PiRi} that $T_c$ is approximately given by the 
well-known BCS relation $T_c \simeq 0.57\Delta$($T$=0). 
In order to derive a similar relation for 
$T_c'$ we inspect the gap equation (\ref{Deltapgap}) at 
$T$~=~0 and in the limit $T \rightarrow T_c'$. Neglecting
$M$ (since $M\ll \mu$ this is valid up to higher
orders in $M^2/\mu^2$) and antiparticle contributions one gets
\begin{alignat}{1}
\int \frac{d^3 p}{(2\pi)^3}
\Big\{&\Big[(1-\frac{{\vec p}_\perp^{\,2}}{s})\frac{1}{E_3^-(\vec p)}
\Big]_{\Delta'(T=0)}
\nonumber\\
   -&(1-\frac{{\vec p}_\perp^{\,2}}{\mu\,|\vec p|})
     \frac{1}{|\mu-|\vec p||} \tanh{\frac{|\mu-|\vec p||}{2T_c'}}\Big\} 
\approx 0~.
\label{Tcest}
\end{alignat}  
Since the integrand is strongly peaked near the Fermi surface,
the $|\vec p|$-integrand must approximately vanish at  $|\vec p| = \mu$,
after the angular integration has been performed. 
From this condition one finds to lowest order in $\Delta'/\mu$:
$  T_c'/\Delta'(T=0) \;\approx\; 1/3~.$
The analogous steps
would lead to $T_c/\Delta$($T$=0) $\approx 1/2$ instead of the
textbook value of 0.57. This gives a rough idea about the quality of the
approximation.    
Note that there are other examples of diquark condensates, where
$T_c \neq 0.57 \Delta$($T$=0)~\cite{SWR02}. This is also the case for
crystalline superconductors~\cite{BKRS01}.

Numerical results for 
$\Delta(T)$ and 
$\Delta'(T)$ are shown in Fig.~\ref{condt}. The quantities have been 
rescaled in order to facilitate a comparison with the above relations
for $T_c$ and $T_c'$. 
Our results
are in reasonable agreement with our estimates. These findings turn
out to be insensitive to the actual choice of parameters.

The specific heat is given by 
$c_v = -T \partial^2 \Omega/\partial T^2$ \footnote{Strictly, 
$c_v = (T/V) (\partial S/\partial T)|_{V,N}$, but
the correction term is small~\cite{FW71}.}.  
For $T \ll T_c$ it is completely dominated by
quarks of color 3, since the contribution of the first two colors
is suppressed by a factor $\rm{e}^{-\Delta/T}$.  Neglecting the
$T$-dependence of $M$ and $\Delta'$, and employing 
Eq.~(\ref{NE}), one finds
\begin{alignat}{1}
  c_v \;\approx\; &\frac{12}{\pi} \frac{\mu^2 +
    |\Delta'|^2}{|\Delta'|}\,T^2\, e^{-\frac{\Delta'_0}{T}}\sum_{n = 0}^3
    \frac{1}{n!}\left(\frac{\Delta'_0}{T}\right)^n~,
\label{cvapp}
\end{alignat}
which should be valid for $T \ll T_c'$. 
In this regime $c_v$ depends quadratically on $T$ for $T \gtrsim \Delta'_0$,  
and is exponentially suppressed at lower temperatures.
\begin{figure}[t]
\epsfig{file=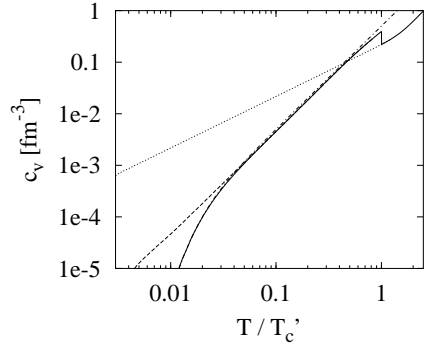,width = 5.8cm}
\caption{Specific heat for parameter set 2 at $\mu$~=~450~MeV as function 
of $T/T_c'$. 
Solid: full calculation,
dashed: result for $M = 0$,
dotted: without spin-1 condensate. The dashed-dotted line indicates the
result of Eq.~(\ref{cvapp}).
}
\label{cv}
\end{figure}
To test this relation we evaluate $c_v(T)$ explicitly using
Eq.~(\ref{Omega}). The results for fixed $\mu$~=450~MeV are displayed
in Fig.~\ref{cv}. For numerical convenience we choose parameter set 2,
leading to a relatively large $\Delta'$($T$=0)~=~30.8~MeV. The
critical temperature is $T_c'\simeq$~0.40~$\Delta'$($T$=0). For the
energy gap we find $\Delta'_0$~=~0.074~$T_c'$.  It turns out that
Eq.~(\ref{cvapp}), evaluated with constant values of $\Delta'$ and
$M$, (dashed-dotted) is in almost perfect agreement with the numerical
result (solid) up to $T \approx T_c'/2$.  The phase transition,
causing the discontinuity of $c_v$ at $T=T_c'$, is of course outside
the range of validity of Eq.~(\ref{cvapp}).  We also display $c_v$ for
$M=0$ (dashed).  Since $\Delta'_0$ vanishes in this case there is no
exponential suppression, and $c_v$ is proportional to $T^2$ down to
arbitrarily low
$T$.  However, even when $M$ is included, the exponential suppression
is partially cancelled by the sum on the r.h.s.  of Eq.~(\ref{cvapp}).
For comparison we also show $c_v$ for a system with $\Delta'$~=~0,
which exhibits a linear $T$ dependence at low
temperatures (dotted).

Our results show that, even though the magnitude of the gap parameter
$\Delta'$ is strongly model dependent 
its relations to the critical temperature and the specific heat are 
quite robust. Thus, if we had empirical data, e.g., for the specific heat
of dense quark matter, they could be used to extract information about the 
existence and the size of $\Delta'$.
In this context neutron stars and their cooling properties are the 
natural candidates to look at.
In Ref.~\cite{RaWi00} it was suggested that the exponential 
suppression of $c_v$ related to the potential pairing of 
quarks of color 3
might have observable consequences for the 
neutrino emission of a neutron star. 
This argument has to be somewhat refined since, as seen above, 
$c_v(T)$ first behaves as $T^2$ and the exponential suppression 
sets in only at 
$T < \Delta'_0$.
The relevance of $c_v$ and the possible effect of diquark condensates
on neutron star cooling was also discussed in Ref.~\cite{neutrino}. 
On the other hand it has recently been argued~\cite{AR02}, that the 
constraints imposed by charge and color neutrality might 
completely prohibit the existence of two-flavor color-superconducting 
matter in neutron stars. 

Because of the spontaneously broken 
$U(1)\times O(3)$ symmetry in Eq.~(\ref{qqt}), for $\Delta' \neq 0$ 
there should be collective Nambu-Goldstone excitations in the spectrum. 
However, due to the Lorentz non-invariance of the system there can be
subtleties \cite{NLSS,HoOM98,MiSh01}.
The NG spectrum can be analyzed within an underlying effective Higgs
potential 
\begin{equation}
V(\phi) = -a^2 \phi_n^\dagger\phi_n 
+ \frac{1}{2}\lambda_1(\phi_n^\dagger\phi_n)^2
+ \frac{1}{2}\lambda_2\phi_n^\dagger\phi_n^\dagger\phi_m\phi_m,
\end{equation}
for the complex order parameter $\phi_n$ \cite{HoOM98}, with 
$\lambda_1 + \lambda_2 > 0$ for stability. 
For $\lambda_2 < 0$ the ground state is 
characterized by $\phi_{vac}^{(1)} = (\frac{a^2}{\lambda_1})^{1/2} (0,0,1)$
which corresponds to our ansatz Eq.~(\ref{qqt}) for the BCS-type
diquark condensate $\delta'$. This solution has the property 
$\langle\vec S\rangle^2 
=(\phi_{vac}^{(1)\dagger} \vec S \phi_{vac}^{(1)})^2 = 0$.
The spectrum of small oscillations above $\phi_{vac}^{(1)}$ consists
of 1+2 NG bosons, all with linear dispersion law: one zero-sound
phonon and two spin waves \cite{HoOM98}. Implying a finite
Landau critical velocity, this fact is crucial for a macroscopic
superfluid behavior of the system \cite{MiSh01}.

Note
that for $\lambda_2 > 0$ there is a
different solution $\phi_{vac}^{(2)} =
(\frac{a^2}{2(\lambda_1+\lambda_2)})^{1/2} (1,i,0)$ with $\langle\vec
S\rangle^2 = 1$.
In this case the NG spectrum above $\phi_{vac}^{(2)}$ 
consists of one phonon with linear dispersion law
and one spin wave whose energy tends to zero with momentum
squared~\cite{HoOM98}. 
The quasiquark dispersion law corresponding to
$\phi_{vac}^{(2)}$ has recently been discussed in Ref.~\cite{ABCC02}. 
A detailed
analysis of the transport properties implied by this type of pairing
seems of particular interest for the phenomenology of neutron stars
since the dispersion law allows for gapless excitations even for
massive quarks~\cite{ABCC02}.

\begin{acknowledgments} 
We thank D. Blaschke, D. Litim, K. Rajagopal, 
I.A. Shovkovy, and E.V. Shuryak for useful discussions.
J.H. thanks J. Wambach and IKP TU Darmstadt for generous 
hospitality and support. 
We acknowledge financial support by ECT$^*$ during its 2001 
collaboration meeting on color superconductivity.
This work was supported in part by grant GACR 202/02/0847.
M.O. acknowledges support from the Alexander von
Humboldt-foundation. 
\end{acknowledgments} 


\begin{thebibliography}{*}
\bibitem{RaWi00} K. Rajagopal and  F. Wilczek, hep-ph/0011333, and references
                 therein.
\bibitem{Al01}   M. Alford, Ann. Rev. Nucl. Part. Sci. {\bf 51}, 131 (2001).
\bibitem{CoPe75} J.C. Collins and M.J. Perry, Phys. Rev. Lett. {\bf 34},
                 1353 (1975).
\bibitem{Son99}  D.T. Son, Phys. Rev. {\bf D59}, 094019 (1999);
                 T. Sch\"afer and F. Wilczek, {\it ibid} {\bf D60}, 
                 114033 (1999);
                 D.K. Hong, V.A. Miransky, I.A. Shovkovy, and 
                 L.C.R. Wijewardhana, {\it ibid} {\bf D61},
                 056001 (2000), {\it err.} {\bf D62}, 059903 (2000).
\bibitem{PiRi}   R.D. Pisarski and D.H. Rischke, Phys. Rev. {\bf D60}, 
                 094013 (1999); {\bf D61} 051501 (2000); 
                 {\bf D61} 074017 (2000).
\bibitem{ARW98}  M. Alford, K. Rajagopal, and F. Wilczek,
                 Phys. Lett. B {\bf 422}, 247 (1998).
\bibitem{RSSV}   R. Rapp, T. Sch\"afer, E.V. Shuryak, M. Velkovsky,
                 Phys. Rev. Lett. {\bf 81}, 53 (1998);
                 Annals Phys. {\bf 280}, 35 (2000).
\bibitem{models} J. Berges and K. Rajagopal, 
                 Nucl. Phys. B {\bf 538}, 215 (1999);
                 G.W. Carter and D. Diakonov,  
                 Phys. Rev. {\bf D60}, 016004 (1999);
                 T.M. Schwarz, S.P. Klevansky, and G. Papp,
                 Phys. Rev. {\bf C60}, 055205 (1999).
\bibitem{BaLo84} D. Bailin and A. Love, Phys. Rep. {\bf 107}, 325 (1984).
\bibitem{RiCD00} D. Rischke, Phys. Rev. {\bf D62}, 034007 (2000);
                 G.W. Carter and D. Diakonov, Nucl.Phys. B {\bf 582}, 
                 571 (2000).
\bibitem{He3}    A.J. Leggett, Rev. Mod. Phys. {\bf 47}, 331 (1975).
\bibitem{Sch00}  T. Sch\"afer, Phys. Rev. {\bf D62}, 094007 (2000).
\bibitem{crystal}M. Alford, J. Bowers, and K. Rajagopal, 
                 Phys. Rev. {\bf D63}, 074016 (2001);
                 R. Rapp, E. Shuryak, and I. Zahed,
                 Phys. Rev. {\bf D63}, 034008 (2001).
\bibitem{ABCC02} M.G. Alford, J.A. Bowers, J.M. Cheyne, and G.A. Cowan,
                 hep-ph/0210106.
\bibitem{SWR03}  A. Schmitt, Q. Wang, D.H. Rischke, nucl-th/0301090.
\bibitem{SWR02}  A. Schmitt, Q. Wang, and D.H. Rischke, 
                 Phys. Rev. {\bf D66}, 114010 (2002).
\bibitem{BKRS01} J.A. Bowers, J. Kundu, K. Rajagopal, and E. Shuster,
                 Phys. Rev. {\bf D64}, 014024 (2001).
\bibitem{FW71}   A.L. Fetter and J.D. Walecka,
                 Quantum theory of many-particle systems, 
                 Mc Graw-Hill, New York (1971).
\bibitem{neutrino} G.W. Carter and S. Reddy, 
                   Phys. Rev. {\bf D62}, 103002 (2000);
                   P. Jaikumar, M. Prakash, T. Sch\"afer, astro-ph/0203088;
                   I.A. Shovkovy, P.J. Ellis, hep-ph/0204132.
\bibitem{AR02}   M. Alford and K. Rajagopal, hep-ph/0204001.
\bibitem{NLSS}   H. Nielsen and S. Chadha, Nucl. Phys. {\bf B105}, 445
                 (1976);
                 H. Leutwyler,  Phys. Rev. {\bf D49}, 3033 (1994);
                 T. Sch\"afer, D.T. Son, M.A. Stephanov, D. Toublan,
                 and J.J. Verbaarschot,
                 Phys. Lett. B {\bf 522}, 67 (2001);
                 F. Sannino and W. Sch\"afer, 
                 Phys. Lett. B {\bf 527}, 142 (2002).
\bibitem{HoOM98} T.-L. Ho, Phys. Rev. Lett. {\bf 81}, 742 (1998);
                 T. Ohmi and K. Machida, J. Phys. Soc. Jpn. {\bf 67},
                 1822 (1998).
\bibitem{MiSh01} V.A. Miransky and I.A. Shovkovy, 
                 Phys. Rev. Lett. {\bf 88}, 111601 (2002).
\end{thebibliography}
\end{document}